\documentclass{PoS}

\usepackage{graphicx}
\usepackage{amsmath,amssymb} 
\usepackage{ulem}

\title{Cluster magnetic fields through the
study of polarized radio halos in the SKA era}

\ShortTitle{Cluster magnetic fields through the study of polarized radio halos in the SKA era}

\author{
\speaker{Federica Govoni}$^1$\thanks{On behalf of the SKA Cosmic Magnetism Working Group} , 
Matteo Murgia$^1$,
Hao Xu$^{2}$,
Hui Li$^{3}$,
Michael Norman$^{2}$,
Luigina Feretti$^{4}$,
Gabriele Giovannini$^{4,5}$,
Valentina Vacca$^{6}$,
Gianni Bernardi$^{7,8}$,
Annalisa Bonafede$^{9}$,
Gianfranco Brunetti$^{4}$,
Ettore Carretti$^{10}$,
Sergio Colafrancesco$^{11}$,
Julius Donnert$^{4}$,
Chiara Ferrari$^{12}$,
Myriam Gitti$^{4,5}$,
Luigi Iapichino$^{13,14}$,
Melanie Johnston-Hollitt$^{15}$,
Roberto Pizzo$^{16}$,
Lawrence Rudnick$^{17}$
\\
$^1$INAF - Osservatorio Astronomico di Cagliari, Italy; 
$^2$University of California at San Diego, CA, USA;
$^3$LANL Los Alamos, NM, USA;
$^4$INAF - Istituto di Radioastronomia Bologna, Italy;
$^5$Dipartimento di Fisica e Astronomia, Universit\'a degli Studi di Bologna,Italy;
$^6$MPA, Garching, Germany;
$^7$SKA SA, Cape Town, South Africa;
$^8$Department of Physics and Electronics, Rhodes University,
Grahamstown, South Africa;
$^9$University of Hamburg, Germany;
$^{10}$CSIRO, Australia;
$^{11}$University of the Witwatersrand, SA;
$^{12}$Observatoire de Nice, France;
$^{13}$Leibniz-Rechenzentrum, Garching, Germany;
$^{14}$Universit\"at Heidelberg, ZAH/ITA, Germany 
$^{15}$Victoria University of Wellington, NZ;
$^{16}$ASTRON, The Netherland;
$^{17}$Minnesota Institute for Astrophysics, University of Minnesota, USA
\\
E-mail: \email{fgovoni at oa-cagliari.inaf.it}
}

\abstract{
Galaxy clusters are unique laboratories to investigate turbulent fluid 
motions and large scale magnetic fields.
Synchrotron radio halos at the center of merging galaxy clusters provide 
the most spectacular and direct evidence of the
presence of relativistic particles and magnetic fields associated 
with the intracluster medium. The study of polarized
emission from radio halos is extremely important to constrain the 
properties of intracluster magnetic fields and the physics of the 
acceleration and transport of the 
relativistic particles. However, detecting this polarized signal is
a very hard task with the current radio facilities.
We use cosmological magneto-hydrodynamical 
simulations to predict the expected polarized 
surface brightness of radio halos at 1.4 GHz.
We compare these expectations with the sensitivity and the 
resolution reachable with the SKA1. This allows us to evaluate 
the potential for studying intracluster magnetic fields  in 
the surveys planned for SKA1.}

\FullConference{
Advancing Astrophysics with the Square Kilometre Array\\
June 8-13, 2014\\
Giardini Naxos, Italy}

\newcommand{\skipthis}[1]{}

\newcommand\apj{ApJ}

\begin{document}

\section{Introduction}
An unambiguous proof of the existence of $\mu$G magnetic fields spread in the 
intracluster medium is confirmed by 
radio observations (Carilli \& Taylor 2002, Govoni \& Feretti 2004).
Information on intracluster magnetic fields can be obtained, 
in conjunction with X-ray observations of the hot gas, 
through the analysis of the rotation measure (RM) of radio galaxies 
lying in the background or embedded within the magnetized 
intracluster medium (Bonafede et al. 2015).
Dedicated software tools and approaches based on a Fourier domain
formulation have been 
developed to constrain the magnetic field power spectrum 
parameters on the basis of the rotation measure 
images (En{\ss}lin \& Vogt 2003, Murgia et al., 2004,
Laing et al. 2008, Kuchar \& En{\ss}lin 2011).
Typically, rotation measure images of cluster radio galaxies permit 
investigating the fluctuations of the intracluster magnetic field below
a spatial scale of about 50-100 kpc.
This does not, however, reach the Mpc scales of 
radio halos (Ferrari et al.2015, Cassano et al. 2015),
which are large-scale diffuse synchrotron sources, located at the center 
of merging galaxy clusters. They have no optical counterpart and no obvious 
connection to cluster galaxies, and are therefore associated with the 
intracluster medium (Feretti et al. 2012).

The study of intracluster magnetic fields is one of the key science 
themes for the SKA Cosmic Magnetism program. The capability of SKA1 to
perform spectro-polarimetric observations with high polarization
purity, angular resolution, and sensitivity will permit us
to dramatically improve our knowledge of the incidence, strength, 
and morphology of large-scale magnetic fields in the Universe.
An All-Sky Polarization Survey at $\simeq$1 GHz (Johnston-Hollitt 
et al. 2015),
will have the potential of measuring the RM toward a large number of 
sources (Krause et al. 2009, Bogdanovi{\'c} et al. 2011)
permitting derivation of a 
detailed description of the strength, structure, and radial decrease 
of cluster magnetic fields.
At the same time, this survey will give the opportunity to investigate 
the total intensity and polarized emission of radio halos at an 
unprecedent sensitivity and resolution, permitting a detailed investigation
of the magnetic field power spectrum in galaxy clusters.
The possibility of obtaining both rotation measure and polarized halo emission
 in the same observation is fundamental to accurately determining the magnetic 
field spectrum over a large range of spatial scales (Govoni et al. 2006).

Cosmological simulations have been playing an
important part in studying the intracluster magnetic field evolution 
of galaxy clusters (e.g. Dolag et al. 2002,
Br{\"u}ggen et al. 2005, Dubois \& Teyssier 2008, Ryu et al. 2008,
Donnert et al. 2009, Xu et al. 2009, Dolag \& Stasyszyn 2009, 
Bonafede et al. 2011, Iapichino \& Br{\"u}ggen 2012).
Although the existence of cluster-wide magnetic fields is now 
well-established, their origin, which is ultimately important for
understanding the evolution of the intracluster medium during the course 
of cluster formation, is still unclear (Widrow 2002). 
Magneto-hydrodynamical simulations of cluster formation have been performed 
with different initial magnetic fields; these have included random 
or uniform fields originating at high redshifts, or from the outflows of normal 
galaxies, or from active galaxies. 
The cluster magnetic fields of all these simulations are roughly 
in agreement at low redshifts. They predict $\mu$G level 
magnetic field strengths in the cluster centers and a decrease of 
the magnetic field strength with radius in the outer regions, in
agreement with the observations. 
Cosmological magneto-hydrodynamical simulations are required at this stage 
of the project to explore the potential of the SKA1 for investigating 
intracluster magnetic fields, especially the possibility of detecting polarized radio halo emission.
This chapter will introduce such simulations and their applications.

\begin{figure*}
\centering
\includegraphics[scale=0.6]{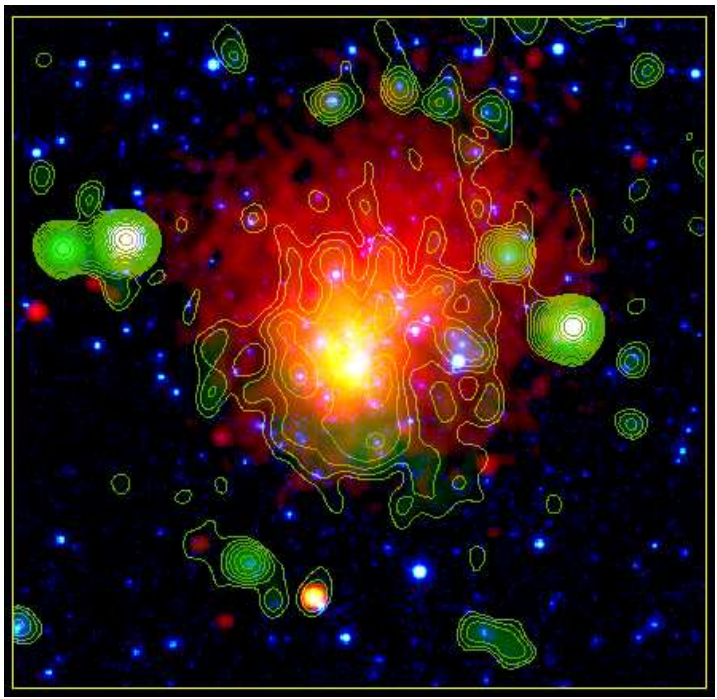}
\caption{
\small
Galaxy cluster A665 at different wavelengths:
optical (blue), X-ray (red), and radio (green contours).
The radio image, showing an extended diffuse radio halo at the
cluster center, has been obtained with the VLA at 1.4 GHz (Vacca et al. 2010).
\normalsize
}
\label{A665}
\end{figure*}

\section{Radio halo studies to investigate cluster magnetic fields}

Sensitive radio observations have revealed diffuse emission from the
central regions of some merging galaxy clusters. These radio sources, 
extending over volumes of $\sim$1 Mpc$^3$ are called radio halos. They are diffuse,
low-surface-brightness ($\simeq$ 1$\mu$Jy~arcsec$^{-2}$ at 1.4 GHz)
and steep-spectrum\footnote{$S(\nu)\propto \nu^{- \alpha}$ structures, with
$\alpha$=spectral index.} ($\alpha>1$) synchrotron sources with
no obvious optical counterparts.

The brightness fluctuations of radio halos are 
closely related to the intracluster magnetic field 
structure and to the spatial distribution of the emitting 
relativistic particles (Tribble 1991). 
Therefore, information on the cluster 
magnetic fields can be derived from detailed radio halo images.
Vacca et al. (2010) presented a study of the 
magnetic field power spectrum in the galaxy cluster A665, which contains 
a Mpc-scale radio halo (see Fig.~\ref{A665}).
Under the assumption that the magnetic field energy density 
is in equipartition with that of the relativistic electrons 
and that these electrons have a power-law energy distribution,
their modeling has proven successful in reproducing the observed total 
intensity fluctuations. 
The magnetic field model that best reproduces the observations 
in A665 is characterized by a central  strength
B$_0$ =1.3$\mu$G, with a magnetic energy 
density decreasing in proportion to the thermal gas density
determined from the intracluster X-ray emission.
Assuming a Kolmogorov magnetic field power spectrum, 
the outer scale of the magnetic field is $\Lambda_{max}$$\sim$450 kpc.

\begin{figure*}
\centering
\includegraphics[scale=0.4]{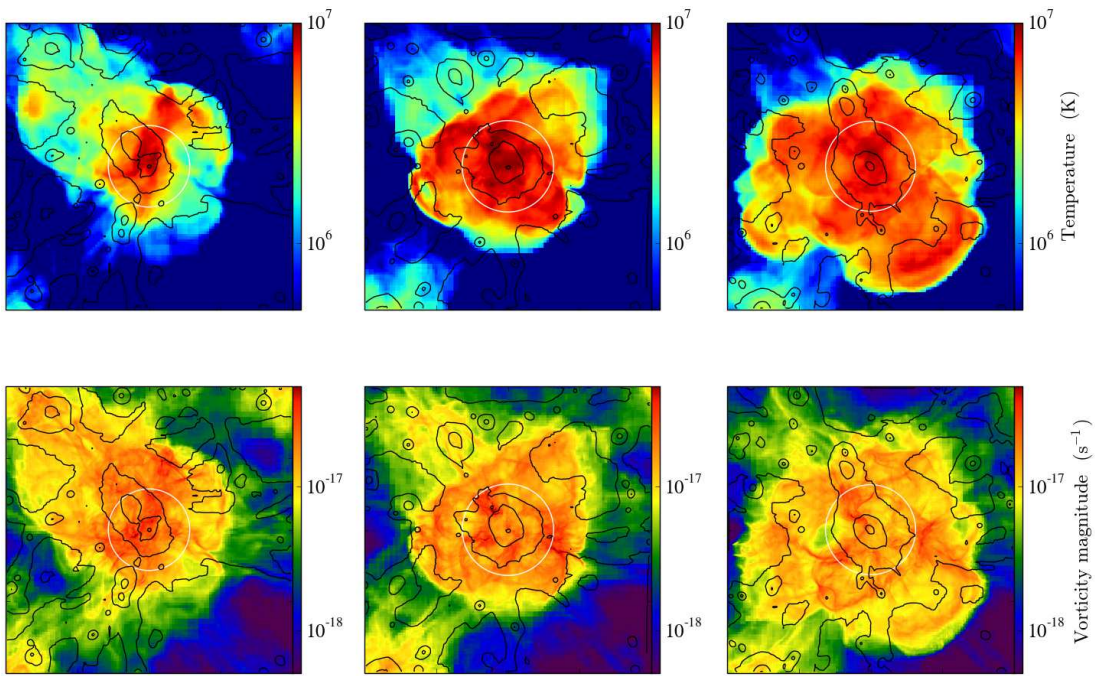}
\caption{
\small
Series of projections ($20\ \mathrm{Mpc}$ on a side, depth of $20\ \mathrm{Mpc}$) 
showing gas temperature (upper row) and vorticity magnitude 
(magnitude of the velocity curl, $|{\bf \nabla} \times {\bf v}|$, lower row)
during a cluster major merger (Iapichino 2014). Three different times are shown, namely pre-merger phase (left, $z=0.43$), post-merger phase (center, $z=0.15$) and later phase (right, $z=0$). The region within one virial radius from the cluster centre is enclosed in the white circle. 
The plots are centered on a massive cluster ($M = 1.3 \times 10^{15}\ M_{\odot}, R_{vir} = 3.1\ \mathrm{Mpc}$ at $z=0$).
\normalsize
}
\label{evolution}
\end{figure*}

Interesting morphological similarities between radio and X-ray
images have been found in a number of clusters hosting a radio 
halo (e.g. Govoni et al. 2004). 
This similarity is generally valid for giant and regular halos. 
However, more irregular and asymmetric halos have been found in the literature.
In these halos, the radio emission may show significant displacement 
from the X-ray emission. We investigated
the statistics of the offset between the radio halo and the X-ray peak of the cluster
emission.
We found that halos can be quite asymmetric with respect to
the X-ray gas distribution, and this becomes more relevant
when halos of smaller size are considered (Feretti et al. 2012,
Govoni et al. 2012). 
A possible explanation for this behavior can be attributed to the
cluster magnetic field power spectrum.
Indeed, on the
basis of their magnetic field modeling, Vacca et al. (2010)
found that if the outer scale of the magnetic field fluctuations is
close to the observing beam, the halo appears smooth and
rounded. Increasing the magnetic field correlation length
results in a heavily distorted radio halo morphology and in
a significant offset of the radio halo peak from the cluster
center. 
Cluster merger events are expected to release a significant 
amount of energy into the intracluster medium.
This energy is injected on large spatial scales, and then turbulent 
cascades may transport energy to smaller scales. 
This process is expected to affect the micro-physics of the 
intracluster medium : particle transport and acceleration, 
heating of the intracluster medium, and magnetic field amplification
(e.g. Roettiger et al. 1999, Ricker \& Sarazin 2001,
Dolag et al. 2005, Vazza et al. 2009, Vazza et al. 2011,
Xu et al. 2010, Xu et al. 2011, Donnert et al. 2013, 
Brunetti \& Jones 2014).
We thus expect a different morphology
in the radio halo structure
between young and old mergers.
Young and smaller systems should have a magnetic field correlation 
length larger than in the more extended, and dynamical older, 
radio halos (Govoni et al. 2012). 
Initially the magnetic field is indeed stirred by the gas-dynamics on scales 
that are larger than that in dynamically older systems. 
In addition, both particle acceleration and diffusion/transport 
depend on turbulent properties (Brunetti \& Jones 2014)
and a smoother spatial distribution of particles is generally 
expected in the case of old radio halos.
In this framework of energy cascades, 
smaller halos may have a more distorted morphology 
if they are young systems in which the energy is still on large scales.
They would have a magnetic field correlation length larger than in
the more extended, and dynamical older, radio halos. 
This could be tested by comparing SKA1 observations 
with the dynamical state of the clusters, 
as seen in the optical and X-ray bands.

The turbulent dynamo is a widely accepted mechanism for amplifying the magnetic field in the intracluster medium to values well above the level predicted by adiabatic compression. This process is likely to also be important  in the cluster outskirts (around and beyond $0.5\ R_{vir}$), where simulations predict a pressure support by turbulent motions of the order of 10--30 per cent of the total 
pressure on scales of a 
few times $100\ {\rm kpc}$ (Vazza et al. 2011). 
Interestingly, recent hydrodynamical 
simulations (Iapichino 2014, Miniati 2014) show that a major merger event can stir the gas and drive turbulent flow even beyond the cluster virial radius.  This could persist up to a few Gyr after the merger, thus providing an important mechanism for amplifying magnetic fields not only in cluster cores, but also in the outer regions (see Fig.~\ref{evolution}).

\subsection{Polarized intensity of radio halos}

The detection of polarized emission from radio halos
would be extremely important to investigate the magnetic field 
power spectrum in galaxy clusters and to find merger shocks in
the intracluster medium not visible in the X-ray images.
Murgia et al. (2004) simulated 3D magnetic 
fields in galaxy clusters with a single
power-law power spectrum of the magnetic field fluctuations 
$|B_k|^2\propto k^{-n}$. 
They investigated how different magnetic field 
power spectra affect the shape and the polarization properties 
of radio halos. Models with $n >3$ and
$\Lambda_{max}$ >100 kpc result in magnetic fields whose energy is larger 
on the large spatial scales, thus giving rise to possible filamentary and 
polarized radio halos. Models with $n < 2$, instead, having most of the 
magnetic field energy on small spatial scales, will give rise to halos 
with a more regular morphology, and very little polarization. 

\begin{figure*}
\centering
\includegraphics[scale=0.7]{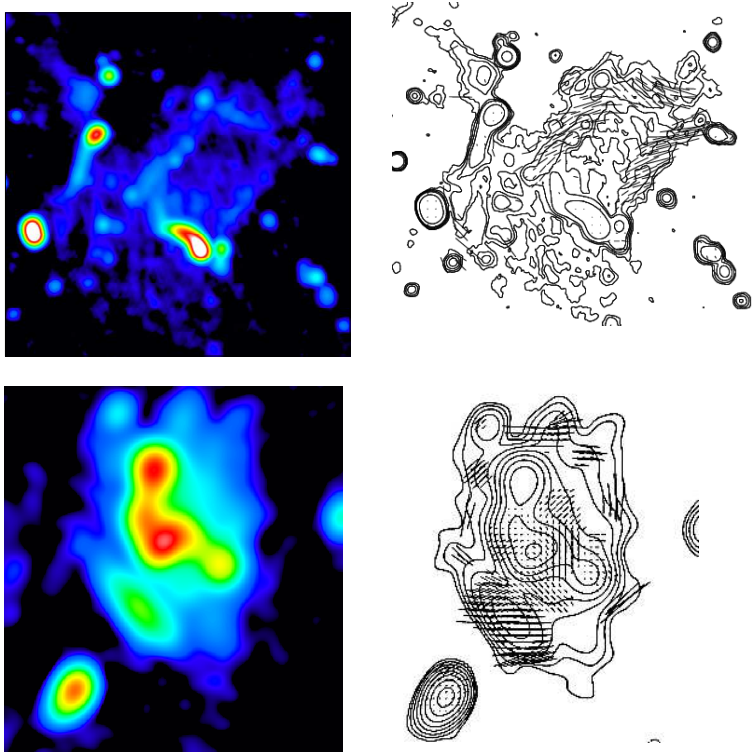}
\caption{
\small
Total intensity and polarization vectors of the radio halos
in which a polarized signal has been detected up to now.  
Top: A2255 (Govoni et al. 2005). Bottom: 
MACS J0717.5+3745 (Bonafede et al. 2009). 
\normalsize
}
\label{pol}
\end{figure*}

Typically, radio halos are found to be unpolarized when observed 
with  current radio facilities.
But, in agreement with the expectations of 
Murgia et al. (2004), filaments of polarized emission
associated with a radio halo have been detected (see Fig.~\ref{pol}) in 
A2255 (Govoni et al. 2005) 
and MACS J0717.5+3745 (Bonafede et al. 2009).
The radio halo in MACS J0717+3745 is one of the 
most powerful radio halos observed so far, with a mean fractional 
polarization of $\simeq$5 \% at 1.4 GHz.
The detection of polarized emission indicates that 
the magnetic field power spectrum in the cluster should be steeper 
than $n=3$, and that its maximum scale $\Lambda_{max}$ should be $>$100 kpc.
On the basis of the interpretation by Govoni 
et al. (2005), in A2255 a single power law cannot account for 
the observed polarization. A power spectrum with spectral index $n = 2$ at the 
cluster center and $n = 4$ at the cluster periphery is needed to 
produce the observed polarized emission. To explain the highly 
polarized structures observed in the radio halo it is necessary 
that most of the magnetic field energy density is on large scales at the cluster
periphery. Pizzo et al. (2011) confirmed that the polarized 
filaments are located at the outskirts of the cluster, but in their 
interpretation such polarization is not associated with the radio halo 
emission.
 
There are several reasons why halo polarization is usually not detected. 
The orientation of the magnetic field likely changes many times
during the path through the cluster.
Faraday rotation along the line of sight will suppress the net
polarization, due to multiple field reversals along each path. Further
depolarization is likely within each beam (beam-depolarization) 
if the beam size
is larger than the angular scale of coherent magnetic field regions.

Recently, Govoni et al. (2013) 
using cosmological magneto-hydrodynamical simulations 
showed that radio halos are intrinsically polarized at full-resolution.
The fractional polarization at the cluster center
is $\simeq 15-35$\% with values varying 
from cluster to cluster and increasing with the distance from
the cluster center. However, the polarized signal is undetectable
if observed with the comparatively shallow sensitivity and
low resolution of current radio interferometers. 

A high level of intrinsic polarization is also qualitatively in agreement with the early post-merger phase in major mergers, when merger shocks are launched in the cluster's innermost region. 
Simulations of major mergers (e.g. Iapichino 2014, Miniati 2014), suggest that, in this early phase, the flow in the core is dominated by the shock propagation, rather than by turbulent stirring. Targeted studies are needed to determine the duration of this transition in comparison with the radio halo lifetime, and the resulting level of polarization.    

\section{Magneto-hydrodynamical simulations} 

\begin{figure*}
\centering
\includegraphics[scale=0.6]{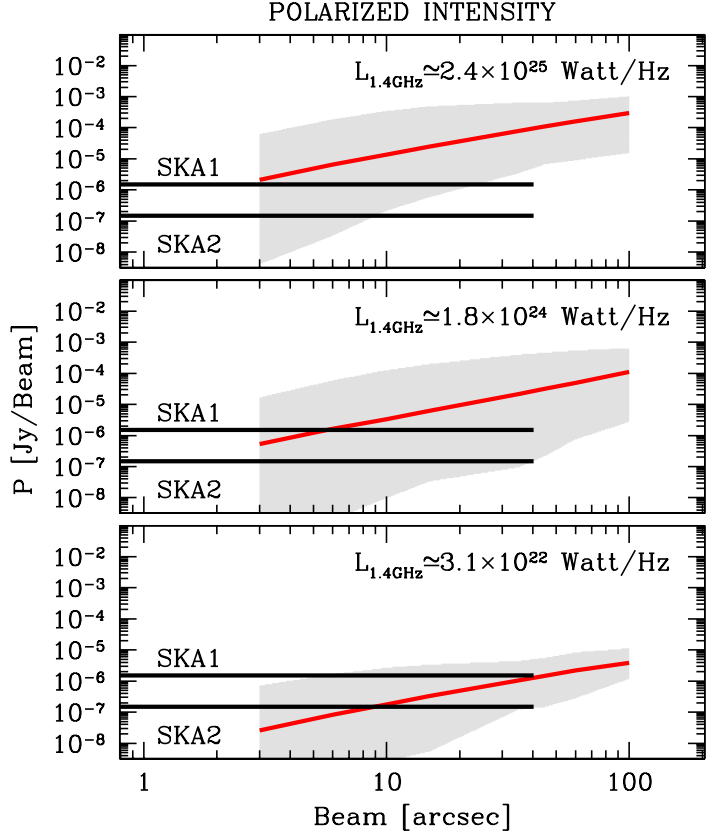}
\caption{
Polarized intensity surface brightness at 1.4GHz as a function of the beam size 
(3$''$-100$''$) for three mock radio halos of different 
luminosities (Govoni et al. 2013), obtained by 
using cosmological magneto-hydrodynamical simulations by Xu et al.
(2012). 
The solid red line shows the average brightness,
while the shaded region shows the maximum and minimum brightness 
fluctuations. 
The simulated surface brightness is compared with a typical sensitivity 
expected for SKA1 ($\sigma \simeq 1\mu$Jy/beam) and 
SKA2 ($\sigma \simeq 0.1\mu$Jy/beam).
The sensitivity refers to the 3-$\sigma$ limit.}
\label{ska}
\end{figure*}

In preparation for SKA, several next-generation radio telescopes 
and upgrades are being constructed around the world.
Among them, APERTIF, ASKAP, LOFAR, Meerkat, and the Jansky VLA 
all offer the chance to explore the polarization properties 
of cluster diffuse emission.
Deep polarization sky surveys are being planned for many of these telescopes.
The WODAN survey (Westerbork Observations of the Deep APERTIF
Northern-Sky; R{\"o}ttgering et al., 2011) will use
APERTIF to explore at 1.4 GHz the northern sky.
This survey will provide a spatial resolution of $\simeq$15$''$ and 
a sensitivity of about 10$\mu$Jy/beam.
A similar performance will be reached in the southern sky with ASKAP
through the polarization survey POSSUM (POlarization Sky Survey of 
the Universe's Magnetism; Gaensler et al. 2010).
These resolution and sensitivity values 
are very promising for detection of
the polarized emission of the most powerful 
($L_{1.4GHz} \simeq 2.4\times10^{25}$ Watt/Hz)
radio halos, while halos of intermediate luminosity 
($L_{1.4GHz}\simeq 1.8\times 10^{24}$ Watt/Hz) will still be 
hardly detectable (Govoni et al. 2013) at this phase.

The high sensitivity and large bandwidths provided by SKA1-MID and
SKA1-SUR can be used to exploit the rotation measure 
synthesis technique and to extract polarized emission at
different Faraday depths (Burn 1966, Brentjens \& de Bruyn 2005). 
Pizzo et al. (2011) showed that this technique is very 
effective  in studying the polarized emission of both the discrete 
sources and the diffuse emission in a comprehensive way.
However, alternative techniques like the 
Faraday synthesis (Bell \& En{\ss}lin 2012),
modeling of the fractional Q and U as a function of
$\lambda^2$ (Farnsworth et al. 2011), and stacking radio 
polarization (Stil \& Keller 2015) are also under investigation 
as ways to best use the capabilities of the SKA and its precursors.
We analyzed (Govoni et al. 2013) how the polarization of radio halos 
will appear at 1.4 GHz, when observed with a bandwidth of 1GHz.
In Fig.~\ref{ska}, we present the expected 
polarized emission of three mock radio halos (characterized by
different radio luminosities) as a function of the angular resolution,
obtained by applying the rotation measure synthesis.
We tentatively represent the 3-$\sigma$ limit
expected with SKA1 ($\sigma \simeq 1\mu$Jy/beam) 
and SKA2 ($\sigma \simeq 0.1\mu$Jy/beam), by 
assuming that the confusion limit is negligible in polarization.
Fig.~\ref{ska} shows that at the sensitivity reachable with the SKA1
it will be possible to detect, at high-resolution 
($\simeq$5-10$''$), polarized emission in radio halos of high 
and intermediate luminosity.
For fainter halos, ($L_{1.4GHz}\simeq 3.1\times10^{22}$ Watt/Hz), the 
detection of the polarized emission at high resolution is very
difficult even for SKA2. Our simulations are based on equipartition 
between relativistic electrons and magnetic fields and assume that 
relativistic particles have a power law energy distribution. 
Simulations including models 
for acceleration/transport will further shed light on this topic.

\section{Conclusions} 
The detection of polarized signal from radio halos is a very hard task 
with current radio facilities. Therefore, this is an
interesting science case for SKA1. 
The detection of polarized emission from radio
halos may be the key for investigating the magnetic field power 
spectrum in galaxy clusters and to find merger shocks in the intracluster 
medium not visible in X-ray images. Indeed, merger shocks 
with Mach numbers $M\simeq1-2$, do not produce enough heating to be visible
in X-ray images. But, through amplification by compression, 
they may be detectable
in polarization (even before they are seen in total intensity)
when observed with SKA1. It is likely that 
thanks to SKA1 we will be able to detect the polarized emission 
in a large number of radio halos, obtaining a
clearer picture of the power spectrum of the magnetic field fluctuations 
over the entire volume of the cluster.

\noindent
Bell, M.~R., \& En{\ss}lin, T.~A., 2012, A\&A, 540, A80 \\

\noindent
Bogdanovi{\'c}, T., Reynolds, C.~S., Massey, R., 2011, ApJ, 731, 7 \\

\noindent
Bonafede, A., Feretti, L, Giovannini, G., et al., 2009, A\&A, 503, 707 \\

\noindent
Bonafede, A., Dolag, 
K., Stasyszyn, F., Murante, G., \& Borgani, S.\ 2011, MNRAS, 418, 2234 \\

\noindent
Bonafede, A., et al., 2015, in proceedings of
"Advancing Astrophysics with the Square Kilometre Array", PoS(AASKA14)095\\

\noindent
Brentjens, M.A., de Bruyn, A.G., 2005, A\&A, 441, 1217 \\

\noindent
Br{\"u}ggen, M., Ruszkowski, M., Simionescu, A., Hoeft, M., Dalla Vecchia, C., 
2005, ApJ, 631, L21 \\ 

\noindent
Brunetti, G., \& Jones, T.W., 2014, International Journal of Modern Physics D, 23, 30007 \\

\noindent
Burn, B.J., 1966, MNRAS, 133, 67 \\

\noindent
Carilli, C.~L., \& Taylor, G.~B.\ 2002, 
Annual Review of Astronomy and Astrophysics, 40, 319\\ 
 
\noindent
Cassano, R., et al., 2015, in proceedings of "Advancing Astrophysics 
with the Square Kilometre Array", PoS(AASKA14)073\\

\noindent
Dolag, K., Bartelmann, M., Lesch, H., 2002, A\&A, 387, 383\\

\noindent
Dolag, K., Vazza, F., 
Brunetti, G., \& Tormen, G., 2005, MNRAS, 364, 753 \\

\noindent
Dolag, K., \& Stasyszyn, F.\ 2009, MNRAS, 398, 1678 \\

\noindent
Donnert, J., Dolag, K., Lesch, H., M{\"u}ller, E. 2009, MNRAS, 392, 1008\\

\noindent
Donnert, J., Dolag, K., 
Brunetti, G., Cassano, R., 2013, MNRAS, 429, 3564\\

\noindent
Dubois, Y., \& Teyssier, R., 2008, A\&A, 482, L13\\

\noindent
En{\ss}lin T.A., \& Vogt C., 2003, A\&A, 401, 835\\ 

\noindent
Farnsworth, D., Rudnick, L., Brown, S., 2011, AJ, 141, 191 \\

\noindent
Feretti, L., Giovannini, G., Govoni, F., \& Murgia, M.\ 2012, The Astronomy and Astrophysics Review, 20, 54\\

\noindent
Ferrari, C., et al., 2015, in proceedings of
"Advancing Astrophysics with the Square Kilometre Array", PoS(AASKA14)075\\

\noindent
Gaensler, B.M., Landecker, T.L., Taylor, A.R., \& POSSUM Collaboration, 
2010, Bulletin of the American 
Astronomical Society, 42, \#470.13  \\

\noindent
Govoni, F., Markevitch, M., Vikhlinin, A., et al.\ 2004, ApJ, 605, 695 \\

\noindent
Govoni, F., \& Feretti, L.\ 2004, International Journal of 
Modern Physics D, 13, 1549\\ 

\noindent
Govoni, F., Murgia, M., Feretti, L., et al., 2005, A\&A, 430, L5 \\

\noindent
Govoni, F., Murgia, M., Feretti, L., et al., 2006, A\&A, 460, 425 \\

\noindent
Govoni, F., Ferrari, C., Feretti, L., et al.\ 2012, A\&A, 545, A74 \\

\noindent
Govoni, F., Murgia, M., Xu, H., et al., 2013, A\&A, 554, A102 \\

\noindent
Iapichino, L., \& Br{\"u}ggen, M.\ 2012, MNRAS, 423, 2781 \\

\noindent
Iapichino, 2014, private communication\\ 

\noindent
Johnston-Hollitt, M., et al., 2015, in proceedings of
"Advancing Astrophysics with the Square Kilometre Array", PoS(AASKA14)092\\

\noindent
Krause, M., Alexander, P., Bolton, R., et al., 2009, MNRAS, 400, 646 \\

\noindent
Kuchar, P., \& En{\ss}lin, T.A., 2011, A\&A 529, A13\\

\noindent
Laing, R.A., Bridle, A.H., Parma, P., Murgia, M., 2008,
MNRAS, 391, 521\\ 

\noindent
Miniati, F.\ 2014, \apj, 782, 21\\

\noindent
Murgia M., Govoni F., Feretti L., et al., 2004,
A\&A, 424, 429\\ 

\noindent
Pizzo, R.F., de Bruyn, A.G., Bernardi, G., Brentjens, M.A., 2011, A\&A, 525, A104 \\

\noindent
Ricker, P.M., \& Sarazin, C.L., 2001, \apj, 561, 621 \\

\noindent
Roettiger, K., Stone, J.M., Burns, J.O., 1999, \apj, 518, 594 \\

\noindent
R{\"o}ttgering, H., Afonso, J., Barthel, P., et al., 2011, 
Journal of Astrophysics and Astronomy, 32, 557 \\

\noindent
Ryu, D., Kang, H., Cho, J., Das, S., 2008, Science, 320, 909\\

\noindent
Stil, J.M., \& Keller, B.W., 2015, in proceedings of "Advancing Astrophysics 
with the Square Kilometre Array", PoS(AASKA14)112\\

\noindent
Tribble, P.C., 1991, MNRAS, 253, 147 \\

\noindent
Vacca, V., Murgia, M., Govoni, F., et al., 2010, A\&A, 514, A71\\

\noindent
Vazza, F., Brunetti, G., Kritsuk, A., et al., 2009, A\&A, 504, 33 \\

\noindent
Vazza, F., Brunetti, G., Gheller, C., Brunino, R., Br{\"u}ggen, M.,
2011, A\&A, 529, A17 \\

\noindent
Widrow, L.~M.\ 2002, Reviews of Modern Physics, 74, 775 \\

\noindent
Xu, H., Li, H., Collins, D.C., Li, S., Norman, M.L., 2009, ApJ, 698, L14\\

\noindent
Xu, H., Li, H., Collins, 
D.C., Li, S., Norman, M.L., 2010, \apj, 725, 2152 \\

\noindent
Xu, H., Li, H., Collins, 
D.C., Li, S., Norman, M.L., 2011, \apj, 739, 77 \\

\noindent
Xu, H., Govoni, F., Murgia, M., et al.\ 2012, \apj, 759, 40 \\

\end{document}